# Performance Evaluation of Deep/Near-Ultraviolet Laser-Assisted Atom Probes for a range of Material system

*Chang-Gi Lee[§], Byeong-Gyu Chae[§], I-Jun Ro, Kyuseon Jang, Eric Woods, Jaemin Ahn, Seong Yong Park, Baptiste Gault, and Se-Ho Kim\**

C.-G. Lee, I.-J. Ro,
Department of Materials Science & Engineering, Korea University, Seoul, 02841 Republic of Korea

B.-G. Chae, J. Ahn, S. Y. Park,
Analytical Engineering Group, Material Research Center, Samsung Advanced Institute of Technology, Samsung Electronics Co., Ltd., Suwon, Republic of Korea

E. Woods
Max Planck Institute for Sustainable Materials (formerly Max-Planck-Institut für Eisenforschung GmbH), Max-Planck-Straße 1, 40237 Düsseldorf, Germany

K. Jang
Department of Materials Science & Engineering, Korea Advanced Institute of Science and Technology, Daejeon, 34141, Republic of Korea

B. Gault
Max Planck Institute for Sustainable Materials (formerly Max-Planck-Institut für Eisenforschung GmbH), Max-Planck-Straße 1, 40237 Düsseldorf, Germany
Department of Materials, Imperial College London, London, SW7 2AZ, United Kingdom

S.-H. Kim\*
Department of Materials Science & Engineering, Korea University, Seoul, 02841 Republic of Korea
Max Planck Institute for Sustainable Materials (formerly Max-Planck-Institut für Eisenforschung GmbH), Max-Planck-Straße 1, 40237 Düsseldorf, Germany
Email: sehonetkr@korea.ac.kr

Keywords: atom probe tomography, normal ultraviolet, deep-ultraviolet, field evaporation

**Abstract**

Atom probe tomography (APT) enables near-atomic-scale three-dimensional elemental mapping through the controlled field-evaporation of surface atoms triggered by the combined application of a DC voltage and either voltage or laser pulses. As the selected laser wavelength for the atom probes transitioned from the near-infrared (1050–1064 nm) to shorter wavelengths e.g., green (532 nm) and near ultraviolet (NUV: 355 nm), the quality of data improved and the range of materials amenable for analysis broadened. A new commercial laser atom probe with a wavelength of 257.5 nm, referred to as deep ultraviolet (DUV), has been recently launched. However, the effects of DUV lasers on different classes of materials have not yet been systematically investigated. In this study, a range of materials, including metals, semiconductor, and oxides, have been examined using commercial atom probes with different laser wavelengths but in principle comparable particle detection system. The quality of the NUV- and DUV-laser atom probe data is evaluated based on four key metrics: background, detection events, ion detection histogram, and mass-resolving power. Furthermore, the application of a thin coating to the finished APT specimens enhances the data quality for both laser wavelengths.

# 1. Introduction

The development of new and innovative materials has been accompanied by advances in microscopy techniques, such that a physicochemical understanding of materials could be gained at the atomic level that can facilitate the development of new materials. It is therefore evident that atomic resolution microscopy is an indispensable tool for the discovery of new structure-property and chemistry-property relationships, which in turn facilitates the development of new materials.

Atom probe tomography (APT) is an advanced material characterization technique that produces sub-nanometer resolved 3D characterization with chemical or elemental identification. The application of an electric potential alongside a voltage or laser pulse to a needle-shaped specimen with an apex diameter of less than 100 nm results in the field evaporation of individual ions, which then fly over a known distance to a position-sensitive 2D detector. The ion time-of-flight, which can be corrected using a reflectron, provides each ion's mass-to-charge ratio.[1–6] Initially, APT used only voltage pulsing to trigger ionization, which largely limited its analysis to conductive materials. More recently, laser pulsing has been introduced, which allowed to diversify the specimen types to include nonconductive glasses,[7] oxides,[8,9] frozen liquids,[10–12] and Li-ion battery electrodes.[13–16]

To date, commercial APT systems have used different pulsed-laser wavelengths, beginning with infrared (1050–1064 nm),[17,18] progressing to green (532 nm),[19,20] and ultraviolet (355 nm).[21–23] Laser irradiation primarily induces thermally-driven field evaporation. Nevertheless, a substantial drawback of pulsed-laser APT is the formation of a heat-affected volume at the apex of the specimen, which can result in unfavorable interdiffusion and asymmetrical shape evolution of APT specimens during the analysis.[17,24,25] Minimizing the heat-affected volume is crucial for enhancing mass-resolving power (MRP), optimizing detection events, and reducing background levels in the mass spectrum. A larger heated volume increases the temperature near the apex, causing an undesirable delayed field evaporation of atoms, degraded MRP, and increased multiple events and background noise in the mass spectrum.[4,24,26,27]

A practical method to mitigate these undesirable effects is to reduce the diameter of the laser spot to create a thermal gradient at the apex of the nanosized APT specimen. This approach to heat dissipation from the heat-affected volume reduces the thermal relaxation time, enabling the specimen to return to the base cryogenic temperature more rapidly and preventing an excessive temperature increase. Consequently, these effects enhanced mass spectral and ion spatial resolutions.[17] Following this development, the local electrode atom probe concept was commercialized, with the first LEAP 3000 system employing a green laser (532 nm),[2,19] and later models, such as the LEAP 4000[28] and 5000 systems,[19,20] used near-ultraviolet lasers (355 nm) (NUV). The latest APT system has adopted a deep-ultraviolet (266 nm) (DUV) laser system on the Invizo and LEAP 6000,[22,29] as it is widely believed that shorter wavelengths can enhance data quality and measurement yield.[30]

The effectiveness of light absorption depends primarily on the absorption coefficient and reflectivity of the material. When materials are exposed to light, free electrons oscillate within the optical field (Drude theory) and gain energy, which is then transferred to the lattice through collisions with phonons and ions, thereby increasing the lattice temperature. The additional ballistic motion of the electrons following energy absorption can lead to an enhanced heated volume.[30] Nevertheless, questions remain regarding the efficiency of this process when the laser wavelength falls below the absorption limit of the material.[26,31]

In the present study, we evaluate the data quality of a DUV-laser system (all measured by Invizo 6000) compared to a NUV-equipped LEAP using various materials, including metals (W, Al, and Fe), semiconductor (Si), and oxides (FeO, SrTiO$_3$, LiCoO$_2$, and LiNi$_{0.8}$Mn$_{0.1}$Co$_{0.1}$O$_2$), along with the use of capping layers by an *in situ* approach inside the focused-ion beam microscope for coating specimens. The data quality of the two different laser sources was differentiated in terms of background, detection event, ion detection histogram, and MRP, using the charge-state ratio (CSR) as a tracer of the electrostatic field to ensure that the data can be compared across datasets and instruments.[32] One limitation of this study is that the Invizo 6000 system has several features that differ from those of the LEAP systems. These include different ion optics with decelerating lenses and DUV-laser illumination from the opposite sides. These differences could introduce additional factors into the analysis that were not considered in the present study. Further, this study emphasized the material-specific outcomes and performance variations observed across different metals, semiconductor, and oxides rather than drawing generalized conclusions. By analyzing these distinct cases under various experimental conditions, the research highlights the complexity and variability inherent in APT measurements, which are influenced by factors such as material properties and the specific APT systems used.

## 2. Results and Discussion

Table 1 shows the experimental conditions for each specimen. We will compare DUV and NUV as a case study under various experimental conditions for each specimen. Table 2 presents a summary of indicative results. While the DUV system demonstrates some minor advantages, these benefits may be attributed to specific characteristics of the specimen rather than the system's inherent superiority. Additionally, there are instances where the DUV setup exhibits slight drawbacks, further suggesting that the overall performance differences are subtle.

**Table 1**. The experimental conditions of each specimen

| Materials | Metal | | | | | | Semiconductor | |
|---|---|---|---|---|---|---|---|---|
| | W | | Fe | | Al | | Si | |
| Beam source | DUV | NUV | DUV | NUV | DUV | NUV | DUV | NUV |
| Temperature (K) | 50 | 60 | 50 | 60 | 50 | 60 | 50 | 50 |
| Pulse rate (kHz) | 200 | 200 | 200 | 200 | 200 | 100 | 200 | 250 |
| Pulse energy (pJ) | 60 | 65 | 50 | 40 | 400 | 20 | 50 | 20 |
| Detection rate (%) | 1 | 0.5 | 2 | 1 | 2 | 1 | 1 | 0.5 |
| Background (ppm/ns) | 18 | 10 | 15 | 3 | 4 | 12 | 6 | 4 |
| Pressure (Torr) | $2.4 \times 10^{-11}$ | $3.0 \times 10^{-11}$ | $7.0 \times 10^{-11}$ | $1.1 \times 10^{-10}$ | $5.7 \times 10^{-11}$ | $2.4 \times 10^{-11}$ | $1.8 \times 10^{-11}$ | $2.7 \times 10^{-11}$ |

| Materials | Oxide | | | | | | | |
|---|---|---|---|---|---|---|---|---|
| | FeO | | $SrTiO_3$ | | $LiCoO_2$ | | $LiNi_{0.8}Co_{0.1}Mn_{0.1}O_2$ | |
| Beam source | DUV | NUV | DUV | NUV | DUV | NUV | DUV | NUV |
| Temperature (K) | 30 | 50 | 50 | 20 | 50 | 60 | 50 | 50 |
| Pulse rate (kHz) | 200 | 200 | 200 | 250 | 200 | 125 | 200 | 100 |
| Pulse energy (pJ) | 20 | 40 | 100 | 200 | 10 | 5 | 1 | 5 |
| Detection rate (%) | 5 | 1 | 1 | 0.3 | 1 | 0.5 | 1 | 1 |
| Background (ppm/ns) | 6 | 7 | 10 | 7 | 40 | 6 | 200 | 10 |
| Pressure (Torr) | $5.8 \times 10^{-11}$ | $1.7 \times 10^{-11}$ | $5 \times 10^{-11}$ | $7.0 \times 10^{-11}$ | $5.5 \times 10^{-11}$ | $6.5 \times 10^{-11}$ | $7.5 \times 10^{-11}$ | $8.6 \times 10^{-11}$ |

| Materials | $LiCoO_2$ | |
|---|---|---|
| | Not coated | Coated by Cr |
| Beam source | DUV | DUV |
| Temperature (K) | 50 | 50 |
| Pulse rate (kHz) | 200 | 200 |
| Pulse energy (pJ) | 10 | 10 |
| Detection rate (%) | 1 | 1 |
| Background (ppm/ns) | 50 | 18 |
| Pressure (Torr) | $2.3 \times 10^{-11}$ | $2.1 \times 10^{-11}$ |

**Table 2.** Comparison charts of DUV-APT and NUV-APT.

| Materials | Metal | | | | | | Semiconductor | |
|---|---|---|---|---|---|---|---|---|
| | W | | Fe | | Al | | Si | |
| Laser source | DUV | NUV | DUV | NUV | DUV | NUV | DUV | NUV |
| Background | < | | < | | > | | ≒ | |
| MRP | > | | < | | > | | < | |
| Ion detection histogram | < | | < | | < | | < | |
| Detection event | < | | < | | > | | < | |

| Materials | Oxide | | | | | | | |
|---|---|---|---|---|---|---|---|---|
| | FeO | | $SrTiO_3$ | | $LiCoO_2$ | | $LiNi_{0.8}Co_{0.1}Mn_{0.1}O_2$ | |
| Laser source | DUV | NUV | DUV | NUV | DUV | NUV | DUV | NUV |
| Background | ≒ | | > | | < | | < | |
| MRP | > | | > | | > | | > | |
| Ion detection histogram | < | | < | | < | | < | |
| Detection event | > | | ≒ | | < | | < | |

| Materials | $LiCoO_2$ | |
|---|---|---|
| | Not coated | Coated by Cr |
| Laser source | DUV | DUV |
| Background | < | |
| MRP | > | |
| Ion detection histogram | < | |
| Detection event | < | |

To assess the data quality of the DUV- and NUV-laser-assisted APT, we conducted a comparative analysis of datasets from a wide variety of materials, including metals (W, Fe, and Al), semiconductor (Si), and oxides (FeO, $SrTiO_3$, $LiCoO_2$, and $LiNi_{0.8}Co_{0.1}Mn_{0.1}O_2$). Evaluations of the 3D atom maps in Figures 1a and 1b suggest that both laser wavelengths are feasible to produce datasets, demonstrating that all materials can be measured using both DUV and NUV lasers.

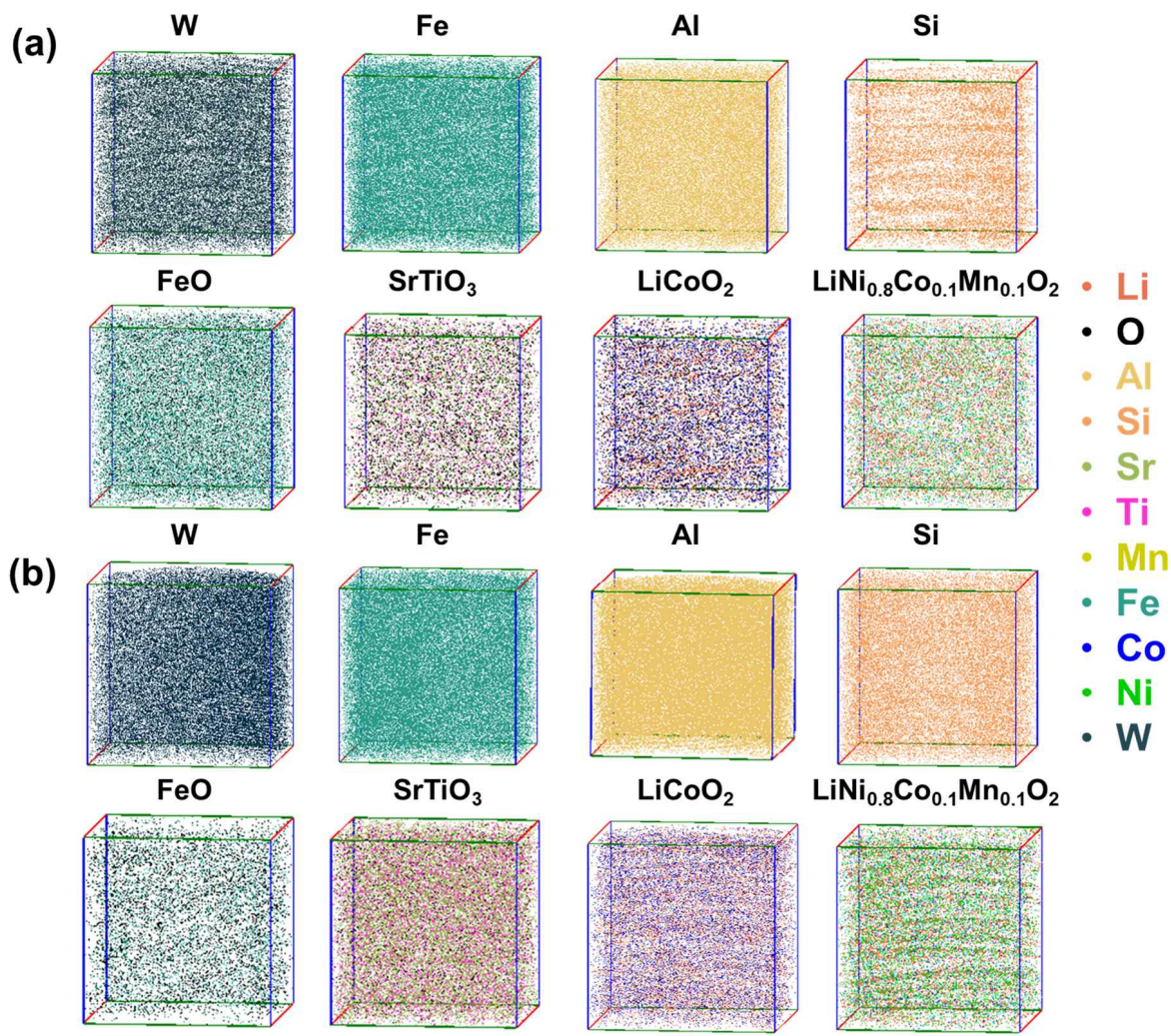

**Figure 1**. 3D reconstruction maps of metals (W, Fe, and Al), semiconductor (Si), oxides (FeO, $SrTiO_3$, $LiCoO_2$, and $LiNi_{0.8}Co_{0.1}Mn_{0.1}O_2$) measured with (a) NUV and (b) DUV lasers. The selected ROI is $10 \times 10 \times 10$ $nm^3$.

## 2.1. Metals and Semiconductors (W, Fe, Al, and Si)

Figure 2a shows the APT mass spectrum of tungsten (W) metal, which is considered a "high evaporation field" metal, with the dominant $W^{3+}$ ion having an evaporation field of 52 V $nm^{-1}$.[33] The CSR for W, defined as the ion count ratio $W^{2+}/W^{3+}$, was used to calculate a DUV-to-NUV metric (DUV/NUV) (e.g., W CSR (DUV)/ W CSR (NUV)), which was 1.25, indicating that a stronger electric field was observed on the W specimen in the DUV measurement than in the NUV measurement (Figure S2a). As the Invizo system has a wider field of view, maintaining a consistent flux across the detector required targeting a higher detection rate (see Table 1) compared to the LEAP system. This adjustment ensures that the flux, when normalized by surface area, remains comparable between the systems.

The background level of approximately 18 ppm ns$^{-1}$ in the DUV measurement was higher than that of approximately 10 ppm ns$^{-1}$ in the NUV (Table 1), as the background level depended on the intensity of the electric field during the analysis.[34] As the electric field increases, the background also increases, and the higher background values in the DUV compared to those in the NUV correspond to this observation.[35] The MRP was calculated by dividing the mass-to-charge ratio of each element (m) by the width of a peak (Δm) at FWHM (i.e. m/Δm). The MRP of DUV at 61.3 ($W^{3+}$) and 92 Da ($W^{2+}$) were 1572 and 1632, respectively, while for NUV, they were 1271 and 1428, indicating enhancement of the MRP of DUV. Figure 2b compares the ion detection histograms of the DUV and NUV, where the DUV tends to show an uneven event contour map. For W, a high-evaporation-field metal, the background, and MRP metrics in the DUV were better than those in the NUV. However, the single-detection efficiency of the DUV (87.9%) was slightly lower than that of the NUV (89.6%). Moreover, the ion detection histogram of DUV showed a concentrated feature at the center, possibly ascribed to the faceting of the specimen at higher temperature or the inherent character of Invizo 6000, which will be discussed.[36]

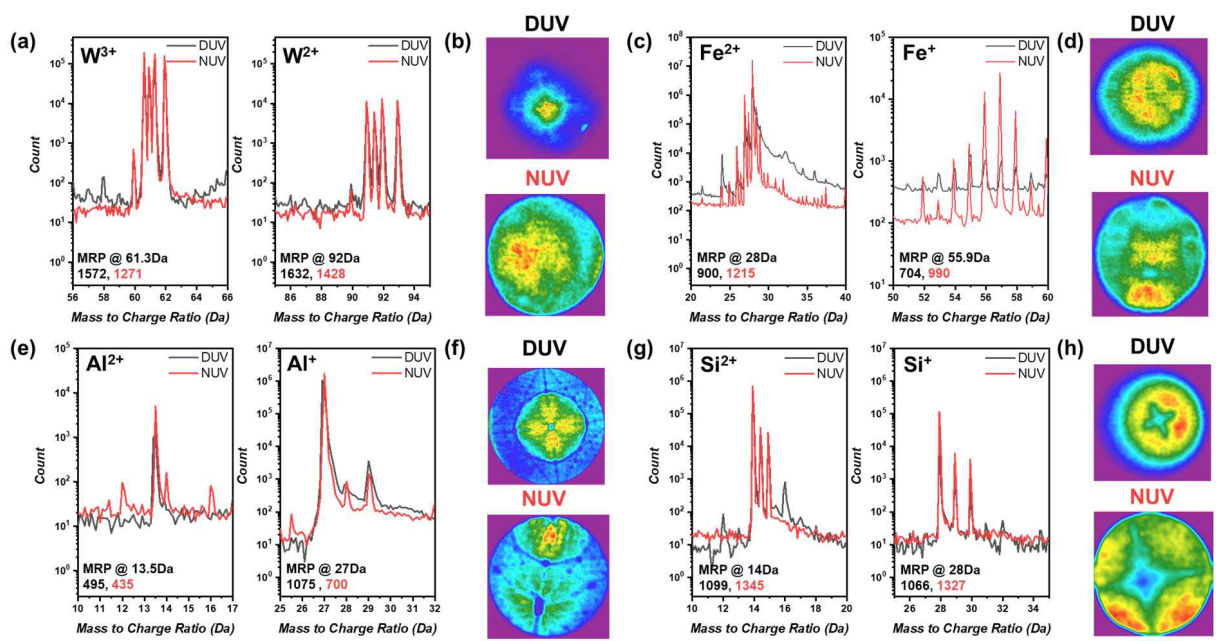

**Figure 2.** Mass spectra and ion detection histograms obtained using DUV and NUV lasers. (a) W, (c) Fe, (e) Al, and (g) Si. The mass resolving power of the DUV and NUV lasers is represented for an indicated peak on each sample. Ion detection histogram of (b) W, (d) Fe, (f) Al, and (h) Si obtained using DUV (upper) and NUV (lower) lasers.

For a "moderate evaporation field" element, low-carbon steel (< 0.03 at.% C) with dominant ion $Fe^{2+}$ having an evaporation field 33 V nm$^{-1}$,[37] was selected and a different trend was observed. As shown in Figure S2a, the DUV/NUV ratio calculated from the $Fe^+/Fe^{2+}$ values was 0.10, indicating that a stronger electric field was applied to the NUV. Despite this, the 15 ppm ns$^{-1}$ background of the DUV was significantly higher than the 3 ppm ns$^{-1}$ background of the NUV, as shown in Table 1. For Fe, NUV showed higher single-detection event rates of 95.3% than 88.0% for DUV, as shown in Table S1. Figure 2c shows that the MRP values for DUV at 28 ($Fe^{2+}$) and 56 ($Fe^+$) Da are 900 and 704, respectively, whereas for NUV, the values at the same Da are 1215 and 990, respectively. Figure 2d shows that the ion detection histogram for the DUV is more concentrated at the center than that for the NUV, which displays an asymmetric pattern affected by trajectory aberrations from the reflectron in the HR series.[38] Overall, for Fe, NUV exhibited better data quality than DUV in terms of detection events, ion detection histogram, MRP, and background.

Finally, examining Al, a "low evaporation field" metal having dominant ion $Al^+$ with evaporation field 19 V nm$^{-1}$ provides additional insights into the differences between DUV and NUV.[33] The DUV/NUV ratio calculated from the $Al^{2+}/Al^+$ counts, is 0.19, implying that a higher electric-field strength was used in the NUV measurement. As expected from those of NUV datasets, the NUV dataset showed a higher background (approximately 12 ppm ns$^{-1}$) than the 4 ppm ns$^{-1}$ background level in the DUV (Table 1). Generally, low-evaporation-field materials tend to have higher background levels. Kim et al. reported on the direct-current (DC) field evaporation of fusible alloys of BiInSn, Ga, and GaIn,[39,40] suggesting that the lack of laser light absorption in low-evaporation-field metals generates continuous field evaporation. The absorption coefficient determines the increase in temperature of the APT specimen tip under laser illumination through thermal activation.[41,42] Al is also considered a low-evaporation-field metal[43,44] and has a relatively low laser absorption coefficient (0.08 for NUV and 0.09 for DUV wavelength).[44] Here, because the observed electrical field strength was high, the lower DUV background for Al could have originated from the slightly increased absorption coefficient in the DUV regime.[45]

Figure 2e shows the MRP values at 13.5 Da ($Al^{2+}$) and 27 Da ($Al^+$) for DUV as 495 and 1075, respectively, whereas NUV shows values of 435 and 700, respectively. The single-event rates were 90.9% for the NUV and 95.0% for the DUV (Table S1). The ion detection histogram for the DUV shows a concentrated density of ions at the center, whereas the NUV shows a uniform histogram across the detector, as shown in Figure 2f.[46] A similar tendency was observed by Tegg et al.,[21] who reported that the ion detection histogram of Al sample on Invizo 6000 showed a high-density region around the center. This histogram pattern could be attributed to a higher temperature at DUV, leading to the faceting of the specimen or inherent system of Invizo 6000 as mentioned above.[47] For Al, the DUV exhibited better results in terms of background, MRP, and detection events, whereas the NUV showed more homogeneous ion detection histogram.

In Figure S2, the DUV/NUV ratio, calculated from the CSR and defined as the $Si^{2+}/Si^+$ value,[48] is 13.6. Despite the DUV having a higher field than the NUV, the background of the DUV at approximately 6 ppm $ns^{-1}$ was similar to that of the NUV at approximately 4 ppm $ns^{-1}$. This different trend can be explained by the higher laser energy of DUV which lowers the applied DC voltage, reducing field evaporation not in time pulse.[49] Figure 2g shows that the MRP values for the DUV at mass-to-charge ratios of 14 ($Si^{2+}$) and 28 ($Si^+$) Da were 1099 and 1066, respectively, whereas the NUV exhibited values of 1345 and 1327, respectively, highlighting the superior mass resolution of the NUV. Moreover, Figure 2h demonstrates that the ion detection histogram of the Si ions for the DUV is compressed, whereas the NUV displayed a more distinct and uniformly distributed pole shape. As shown in Table S1, the single-detection event rates were 91.0% for the DUV and 96.26% for the NUV, suggesting a marginally higher detection efficiency for the NUV. Overall, for Si, DUV demonstrated a similar background and improved results in terms of single-detection event; however, it underperformed not only in the MRP but also in the ion detection histogram, which is directly linked to the 3D reconstructed atom map, compared to NUV.

It is important to note that multiple parameters influence these measurements, making it challenging to isolate the specific effect of the laser on the observed differences. Significant efforts were made to control other variables, such as maintaining consistent sample preparation and measurement conditions, to focus on the laser's impact. Although Invizo 6000 has implemented a dual DUV-laser system and a decelerator to increase data volumes and to improve specimen yields by reducing the electrostatic field stress, particularly targeting semiconductor materials that require uniform heat deposition to trigger field evaporation, the acquired results suggest that DUV does not provide significant benefits for some metals, particularly Fe, compared to NUV-assisted APT. These findings suggest the need for a balanced consideration of both techniques depending on the specific analytical requirements.

### 2.2. Oxides

*2.2.1. Wüstite (FeO)*

FeO was measured using both NUV and DUV, revealing a DUV/NUV ratio of 1.38, calculated from the CSR, defined as the $Fe^+/Fe^{2+}$ ratio, demonstrating that a higher electric field was observed in the DUV measurement (Figure S2b). The background level in the DUV (7 ppm $ns^{-1}$) was similar to that in the NUV (6 ppm $ns^{-1}$). Also, in Table S1, the single-detection event rate for the DUV was 53.7%, whereas that for the NUV was 50.7%. A higher electric field generally increases not only the background but also the multiplicity, which is attributed to the greater possibility of correlative evaporation near neighboring atoms at higher electric fields and unwanted field evaporation at non-pulsing by a high DC voltage. However, the DUV manifested a background similar to that of the NUV and a better multiplicity, albeit with a high electric field. This similar background may be due to the higher temperature of NUV than DUV.[49] In addition, the mesh between the specimen and detector in Invizo 6000 may be attributed to a decrease in multiple events by filtering out the daughter ions that dissociated from the field-evaporated molecule.[38]

DUV also demonstrates an outstanding performance in terms of mass resolution. The MRP values for the DUV at 28 ($Fe^{2+}$) and 56 ($Fe^+$) Da were 718 and 758, respectively, whereas those for the NUV were 571 and 327, respectively (Figure 3a). This indicates that DUV achieves sharper and more distinct peaks, which is crucial for the accurate identification and quantification of different ion species. The NUV peak at 56 Da ($Fe^+$) exhibits a severe thermal tail, which is often present in laser-pulsed APT data for nonconductive materials. These tails are generally attributed to thermal bulk heating and slow cooling mechanisms, resulting in time-delayed thermal desorption of surface ions.[26,42] The superior peak resolution in DUV can be explained by a smaller heat-affected zone (<2 μm), which minimizes peak broadening and overlapping, thereby enhancing the overall resolution of the mass spectrum.[17,50,51] Another possible explanation is that the dual-beam system of the Invizo 6000 reduces the localized laser absorption in one direction, thereby suppressing the time shift caused by delayed heat transfer.

Figure 3b depicts the detection histogram; interestingly, the DUV exhibits a hollow core with a dense outer layer of ion detection, unlike the NUV. A local increase in the electric field corresponding to a region of higher curvature (i.e., in the vicinity of the poles and zone lines) induces local magnification, leading to a region of low density that appears dark in the detector map. The sample used was a (100)-grown single-crystal wüstite (FeO), and we expected a low-density region at the (100) pole. However, the zones of low-density regions are abnormally large, and the field of view is progressively compressed compared to other oxide zones in previous reports.[52,53] This is likely due to trajectory aberrations caused by the field lens, resulting in local inhomogeneities not from the sample, but from the instrument itself.

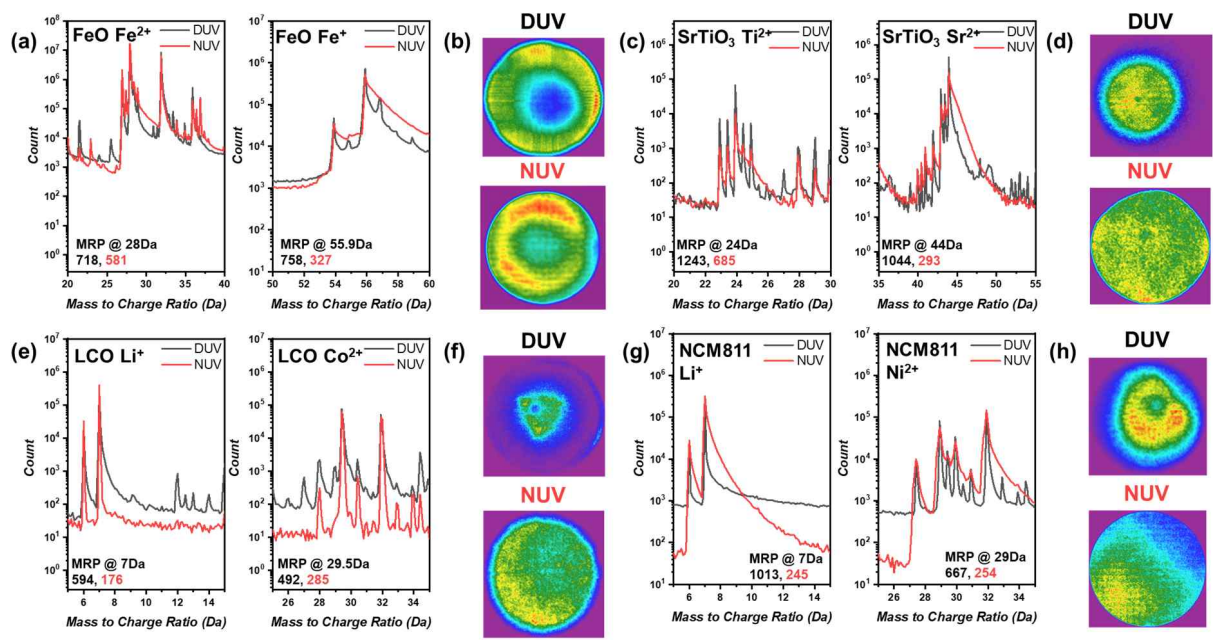

**Figure 3.** Mass spectra and detection histograms obtained using DUV and NUV lasers. (a) FeO, (c) SrTiO$_3$, (e) LCO, (g) NCM811. The mass resolving power of the DUV and NUV lasers is represented for a indicated peak on each sample. Detection histograms of (b) FeO, (d) SrTiO$_3$, (f) LCO, (h) NCM811 obtained using DUV (upper) and NUV (lower) lasers.

*2.2.2. Strontium Titanate (SrTiO$_3$)*

Strontium titanate (SrTiO$_3$) poses significant challenges for measurements using an electrostatic-field-induced evaporation technique because of its intrinsic brittleness and piezoelectric properties, which can cause substantial volume changes when an external voltage is applied. To mitigate the piezoelectric effect, the extrinsic properties of SrTiO$_3$ can be adjusted by introducing coating layers;[9,54] however, APT measurements are still unstable. Despite these challenges, a thin film of SrTiO$_3$ has been analyzed using NUV and DUV-laser-assisted APT because of its high absorption coefficient for both laser systems.[55] The DUV measurements at the 24 (Ti$^+$) and 44 Da (Sr$^{2+}$) mass positions yielded MRPs of 1243 and 1044, respectively, demonstrating a significantly superior mass resolution, as shown in Figure 3c. The DUV/NUV ratio, derived from the CSR and defined as the Ti$^+$/Ti$^{2+}$ ratio, was 3.57, indicating that a higher electric field was observed in the DUV measurements. The background level in the DUV (10 ppm ns$^{-1}$) was slightly higher to the NUV (7 ppm ns$^{-1}$) as expected. The detection histogram in the NUV displays a homogeneous distribution of detected ions, while the DUV shows concentrated ion detection maps in Figure 3d, which can distort the 3D reconstruction map. Overall, for SrTiO$_3$, DUV demonstrated superior performance in terms of MRP while maintaining background levels comparable to those of NUV.

*2.2.3. Metal–Oxide Cathodes*

The analysis of Li-ion battery materials using APT presents significant challenges owing to the in situ delithiation process (i.e. electromigration) by intense electric field during APT measurement and the large evaporation-field difference between Li and transition-metal oxides such as LiCoO$_2$ (LCO) and LiNi$_{0.8}$Co$_{0.1}$Mn$_{0.1}$O$_2$ (NCM811).[14,56,57] Despite those challenges, the battery cathode has been analyzed by optimizing the analysis conditions such as the temperature, pulse rate etc.[14,58] For LCO, the DUV/NUV ratio calculated from the selected CSR Co$^+$/Co$^{2+}$ values was 1.09, indicating that a higher electric field was applied to the DUV measurement. Although the MRP value for the DUV was higher than that of the NUV, the background level in the NUV (6 ppm ns$^{-1}$) was considerably lower than that in the DUV (40 ppm ns$^{-1}$). This indicates a superior peak resolution for DUV, but a loss of ions, as shown in Figure 3e. Moreover, the single-detection event rate for DUV was 63.6% compared to 70.4% for NUV, suggesting a higher detection efficiency for NUV. A similar trend was observed for LiNi$_{0.8}$Co$_{0.1}$Mn$_{0.1}$O$_2$ (NCM811). The peak resolution of the DUV was better than that of the NUV at 7 and 29 Da, as shown in Figure 3g. The CSR value calculated using Ni$^{2+}$/Ni$^+$ was 0.94, and the background in the NUV at approximately 10 ppm ns$^{-1}$ was significantly lower than that in the DUV at approximately 200 ppm ns$^{-1}$. The single-detection rate of the NUV was higher than that of the DUV, analogous to that of the LCO in Table S1. Both DUV results for the battery sample show an ion detection histogram (Figures 3f and 3h) with ions concentrated at the center and formed in a triangular shape, indicating a potential issue with ion distribution. Overall, while the DUV demonstrates a better MRP, the NUV performs better in terms of background, detection events, and ion detection histogram distribution.

The background, detection events, ion detection histograms, and MRP for each material were examined to compare the data quality between the DUV and NUV. In general, the differences between the DUV and NUV depend on materials. For MRP, DUV typically provides better performance, except for certain materials such as Fe and Si. However, the DUV detection histograms consistently showed a centered appearance, suggesting a potential distortion or uneven field distribution.

## 2.3. Application of *In Situ* Coating Methods

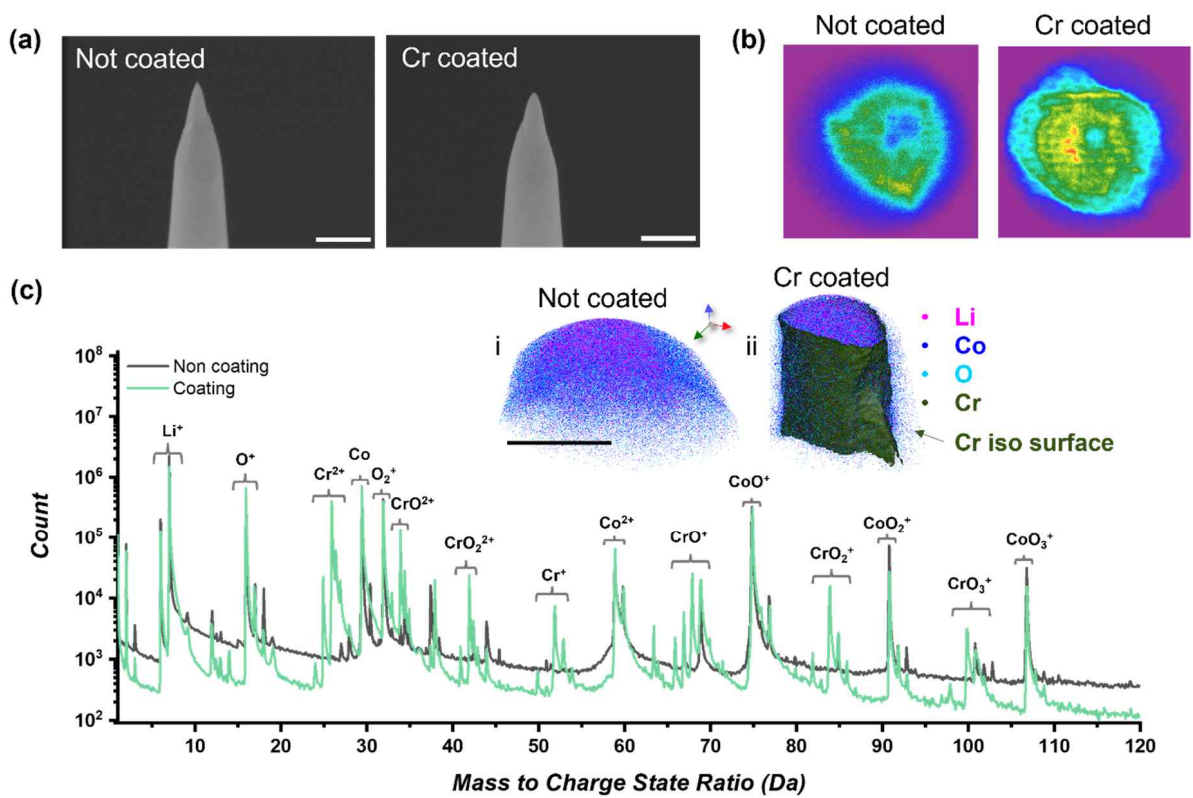

**Figure 4.** SEM images and mass spectra. (a) SEM images captured before and after coating Cr on the LCO specimen. The white scale bar implies 250 nm. (b) Ion histograms of uncoated and coated LCO. (c) Mass spectra uncoated and coated LCO. The inset figures (i) and (ii) are 3D reconstruction maps of LCO and Cr-coated LCO. The black scale bar implies 200 nm.

The LCO samples were coated *in situ* using the focused-ion beam (FIB) method by following the protocols described by Woods et al., Singh et al., and Johansen et al. [59–61] Briefly, a small Cr lamella was mounted on the *in situ* micromanipulator, a half-circle was cut, and Cr was back-sputtered using a Ga-ion beam onto the final sharpened specimen using a low ion-beam current and time. Both Cr- and uncoated LCO samples were analyzed using DUV-APT (Figure 4), and the scanning electron microscopy (SEM) image in Figure 4a shows that the coated sample was blunter than the uncoated sample after *in situ* Cr coating. *In situ* Cr coating helps protect the region of interest (ROI) and improves the measurement quality by stabilizing the material and reducing preferential evaporation.[59] Additionally, it may alter the charge density distribution under electric fields, making nonconductive materials measurable for NUV-APT. By providing a uniform and protective layer directly on the specimen, the coating stabilized the material and reduced background noise, thus enhancing the data quality and yield.[62,63] Furthermore, this coating improves the mass resolution and expands the effective field of view, allowing the visualization of the entire original specimen, including the surface oxide layers.[59,60,64]

The CSR ratio of the coated specimen to the uncoated specimen, calculated from the $Co^+/Co^{2+}$ counts, was 0.743, indicating that the coated samples experienced stronger electrical fields than the uncoated samples. As shown in Figure 4c, the background level of the coated samples was significantly lower than that of the uncoated samples, despite the higher electric field applied to the coated specimens. This improvement in background noise is likely related to thermal dissipation from the sample, as previous studies have reported that thin metal coatings (applied *ex situ* and *in situ*) can improve background noise levels in specimens with poor thermal conductivity.[63,65] At this point, we cannot fully explain why the triangular shape pattern appeared in the ion detection histogram of the uncoated specimen (Figure 4b), reducing the resolution of the measurement in Invizo 6000. After coating, the ion detection histogram for the coated LCO samples showed improvement, displaying suppressed abnormal triangular contour maps and low-density fluctuations at the center of the detector.

In summary, the data quality of analyzed materials using NUV and DUV lasers was thoroughly compared using four metrics (background, MRP, detection events, and ion detection histograms). The results for the background and detection events between NUV and DUV across different material types such as metals, semiconductor, and oxides do not show a distinctive trend in the performance. However, DUV appears to have better MRP than NUV across most metals and oxides. Possible explanations for the improved MRP with DUV are its shorter wavelength, which results in a smaller heat zone and easier thermal dissipation, and the Invizo6000's deceleration lenses for prolonged time-of-flight.

The ion detection histograms on the detector for all samples displayed severely distorted and compressed images, which was likely due to Invizo's newly implemented decelerating lens. When field-evaporated ions pass through the counter electrode toward the detector, a series of decelerating lenses could change the trajectory of these ions. Another possible reason for the concentrated histogram may be the dual-beam source of Invizo 6000. The laser absorption mechanism from dual-beam becomes quite complex due to various effects such as interference caused by interaction between both pulses.[30,66] This complexity contributes to the distortion of the ion detection histogram but it should be investigated further. That concentrated map eventually causes composition inaccuracy and lowers the spatial resolution since the position-sensitive detector cannot distinguish multiple ions nearly simultaneously impacting close spatial proximity on the MCP. [38,67] Furthermore, Artifacts in the 3D reconstruction atom map underestimate the atom distribution due to the high image compression factor.[21,68] Therefore, appropriate reconstruction methods tailored to DUV-APT are required to avoid such distortions.

When bare and *in situ* Cr-coated LCO were measured using DUV, there was a significant improvement in the background and detection events for the coated LCO compared to the uncoated LCO. This suggests that the introduction of the coating process can enhance data quality in DUV-APT systems. In particular, Cr *in situ* coating has emerged as a promising candidate for improving APT measurements, providing a strategy not only to enhance the data quality of both systems but also to reduce the variability between DUV and NUV results.

## 3. Conclusion

DUV- and NUV-laser sources equipped with Invizo 6000, LEAP 4000, and LEAP 5000 were compared across different types of materials (metals, semiconductors, and oxides). In addition, the results of the *in situ* Cr coating are demonstrated in the DUV. Overall, the yields and mass resolution from DUV were better, but no distinct improvement in data quality was observed considering the severely distorted ion histogram map. This discrepancy may be attributed to the differences in the laser wavelength, dual-illumination laser optics scheme, and decelerating lenses. Therefore, a reassessment of the application of laser sources is necessary to improve data quality. Furthermore, applying a coating to the specimen could be an effective method for stabilizing APT specimens from the beam, improving data quality, and narrowing the gap between the two laser sources. These results provide valuable insights for future studies.

## 4. Experimental Section

*Materials*: In this study, metals (W 97.0%, low-carbon Fe alloy 98.0%, and Al 99.9%),

semiconductors (Si 99.5%), and oxides ((100)-orientated single-crystal FeO 99.9%, $SrTiO_3$ 99.9%, $LiCoO_2$, and $LiNi_{0.8}Mn_{0.1}Co_{0.1}O_2$) were used as the primary materials. The metals used for coating (Cr 99.99%) were purchased from commercial sources.

*Characterization*: The samples were sharpened into needles by FIB (Thermo Scientific Helios 5 HX) milling with Ga ions using a combined FIB/SEM instrument. The samples were prepared using either the "toblerone" or standard TEM-lamella lift-out methods, and then mounted on a sharpened wire or pre-sharpened Si microtips with a composite (e.g., Pt/C). This composite was deposited by injecting a gaseous precursor on the FIB/SEM system using a gas injection system.[69,70] The mounted samples were sharpened to a final radius of less than 100 nm by annular milling with a 30-kV Ga-ion beam. In this protocol, a circle was used, in which both the inner diameter and ion-beam current were progressively reduced until the final tip radius was obtained. The sharpened specimens were cleaned with a low-current 5-kV ion beam to eliminate residual Ga-ion-induced beam damage and implantation.

*NUV- and DUV-laser measurements*: The NUV-laser measurements were performed using CAMECA Instruments LEAP4000XR, LEAP5000XR, and LEAP5000XS, whereas for the DUV-laser measurements, an Invizo6000 system (CAMECA Instruments, Madison, WI, USA) was used. The detailed measurement conditions and parameters are listed in Table 1. All the APT data for the metals, semiconductor, oxides, and application sections were reconstructed using the CAMECA Atom Probe Suite (AP Suite) version 6.3, using previously described standard protocols.[71] The samples and equipment models used in this study are listed in Table 3.

**Table 3.** Samples and equipment models.

| Materials | Metal | | | | | | Semiconductor | |
|---|---|---|---|---|---|---|---|---|
| | W | | Fe | | Al | | Si | |
| Wavelength | DUV | NUV | DUV | NUV | DUV | NUV | DUV | NUV |
| Equipment | Invizo 6000 | LEAP 5000XR | Invizo 6000 | LEAP 5000XR | Invizo 6000 | LEAP 5000XS | Invizo 6000 | LEAP 5000XR |

| Materials | Oxide | | | | | | | |
|---|---|---|---|---|---|---|---|---|
| | FeO | | $SrTiO_3$ | | $LiCoO_2$ | | $LiNi_{0.8}Co_{0.1}Mn_{0.1}O_2$ | |
| Wavelength | DUV | NUV | DUV | NUV | DUV | NUV | DUV | NUV |
| Equipment | Invizo 6000 | LEAP 5000XS | Invizo 6000 | LEAP 4000HR | Invizo 6000 | LEAP 5000XR | Invizo 6000 | LEAP 5000XS |

| Materials | $LiCoO_2$ | |
|---|---|---|
| | Not coated | Coated by Cr |
| Wavelength | DUV | DUV |
| Equipment | Invizo 6000 | Invizo 6000 |

*Data Analysis*: Datasets containing the same number of ions were selected for each material (Figure S1) and evaluated using the following metrics: (1) detection events, (2) ion histogram uniformity, (3) MRP, and (4) background levels. Each parameter affects the quality of the APT data as described below:

(1) Multiple events, dissociated ions, or neutrals during the field-evaporation trajectory result in ambiguous spatial resolutions, hindering accurate reconstruction and causing inaccuracies in the composition analysis owing to the loss of ions.

(2) The MRP is related to the chemical resolution; that is, the higher the resolution, the higher the sensitivity and spectral resolution. A low MRP is attributed to the presence of thermal tails in the mass peaks, which lead to a loss of elemental information because the elemental peaks may become indistinguishable below these tails.

(3) Uniformity in the detector histogram is crucial for accurate reconstruction of the 3D atomic structure of an APT specimen. Irregularities originating from sources other the sample can lead to inaccuracies in the spatial resolution.

(4) The background level determines the chemical sensitivity of the measurements. Background noise can be caused by vacuum conditions, hydrogen dissociation in the analysis vacuum chamber, changes in the electric-field strength, and delayed evaporation.[59] The ions evaporated during the pulses exhibit a high kinetic energy and are identified in the mass spectrum. However, the atoms on the specimen surface can also evaporate in-between the applied pulse durations, if the field strength is sufficiently high, and trigger evaporation. The kinetic energy of these atoms may not be sufficient to trigger the MCP detector, and such atoms are identified as background noise events.

In this study, the vacuum pressure did not affect the background because all the measurements were conducted at pressures below $10^{-9}$ mbar. Hydrogen is assumed to be the predominant residual gas in metal vacuum chambers at extremely low pressures. Residual gas molecules can be adsorbed by the tip and ionized by the local electric field at the specimen's surface edge (a process known as "field ionization") and accelerate toward the detector, triggering DC evaporation and creating extra signals in the MCP. In a previous study, the adsorption of residual gases under ultrahigh-vacuum (UHV) conditions was analyzed by combining the equations of classical gas kinetics theory with the electric-field enhancement factor.[72] These calculations revealed that under UHV, approximately $10^4$ years would be required to achieve monolayer coverage of hydrogen molecules on a typical APT specimen for the field-induced ion evaporation process. Because all the experiments were conducted under UHV conditions, we assumed that residual gas ionization influenced only the electric-field strength; it did not cause background noise.

**Supporting Information**

Supporting Information is available from the Wiley Online Library.

**Acknowledgments**

We thank Uwe Tezins, Christian Broß and Andreas Sturm for their support to the FIB and APT facilities at MPIE. S.-H.K acknowledge supports from KIAT-MOTIE (P0023676, HRD Program for Industrial Innovation and RS-2024-00431836, Technology Innovation Program). B.G. acknowledge financial support from the ERC-CoG-SHINE-771602.

# Supporting Information

## Performance Evaluation of Deep/Normal Ultraviolet Laser-Assisted Atom Probes Across Diverse Materials

*Chang-Gi Lee[§], Byeong-Gyu Chae[§], I-Jun Ro, Kyuseon Jang, Eric Woods, Jaemin Ahn, Seong Yong Park, Baptiste Gault, and Se-Ho Kim\**

C.-G. Lee, I.-J. Ro,

Department of Materials Science & Engineering, Korea University, Seoul, 02841 Republic of Korea

B.-G. Chae, J. Ahn, S. Y. Park,

Analytical Engineering Group, Material Research Center, Samsung Advanced Institute of Technology, Samsung Electronics Co., Ltd., Suwon, Republic of Korea

E. Woods

Max-Planck-Institut für Eisenforschung GmbH, Max-Planck-Straße 1, 40237 Düsseldorf, Germany

K. Jang

Department of Materials Science & Engineering, Korea Advanced Institute of Science and Technology, Daejeon, 34141, Republic of Korea

B. Gault

Max-Planck-Institut für Eisenforschung GmbH, Max-Planck-Straße 1, 40237 Düsseldorf, Germany

Faculty of Engineering, Department of Materials, Imperial College London, London, SW7 2AZ, United Kingdom

S.-H. Kim\*

Department of Materials Science & Engineering, Korea University, Seoul, 02841 Republic of Korea

Max-Planck-Institut für Eisenforschung GmbH, Max-Planck-Straße 1, 40237 Düsseldorf, Germany

Email: sehonetkr@korea.ac.kr

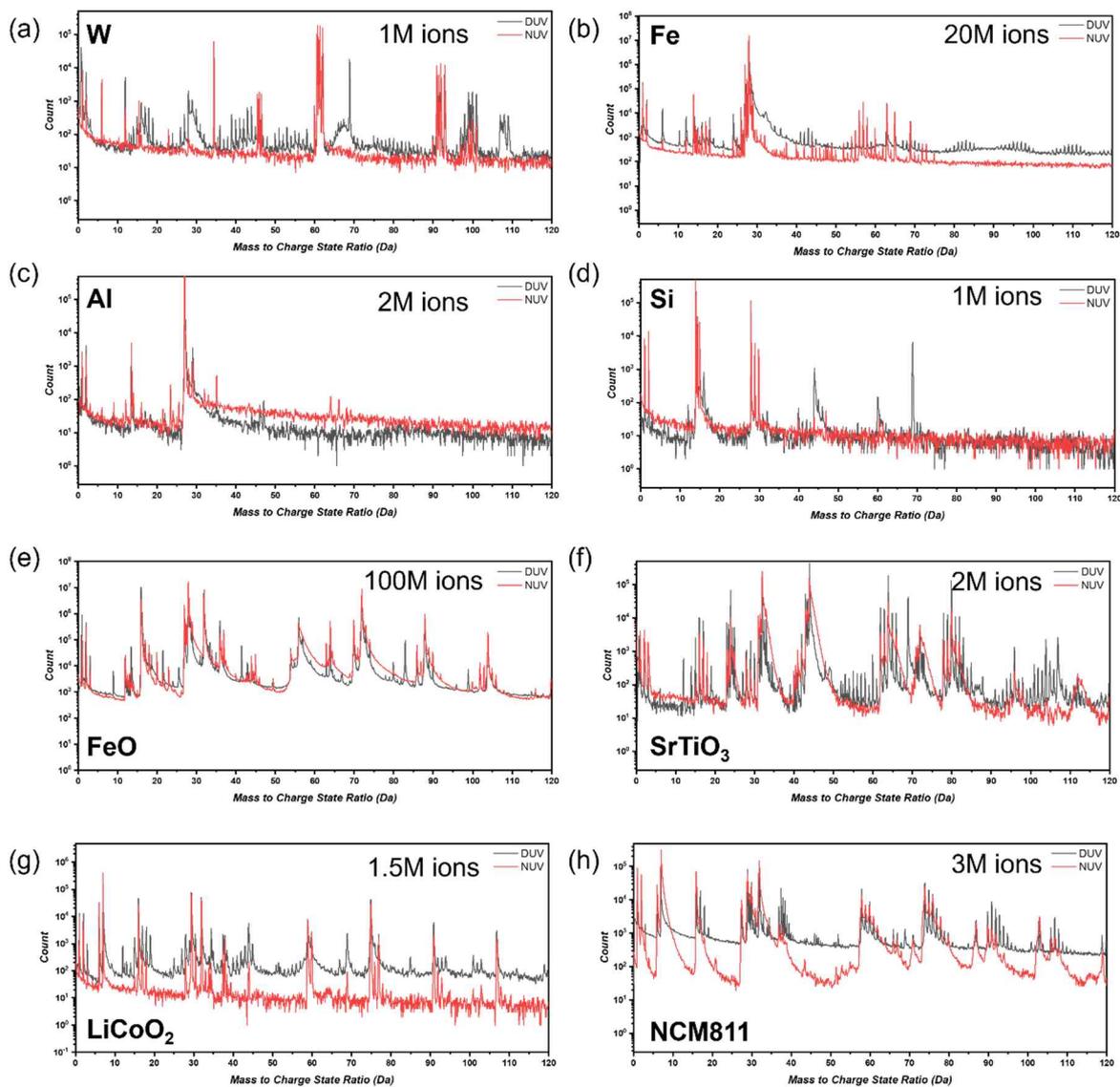

Figure S1. (a-h) Mass spectra of each element, W, Fe, Al, Si, FeO, SrTiO$_3$, LiCoO$_2$, LiNi$_{0.8}$Co$_{0.1}$Mn$_{0.1}$O$_2$ (NCM811). The ion extracted for each spectrum corresponds to 1M, 20M, 2M, 1M, 100M, 2M, 1.5M, and 3M, respectively.

.

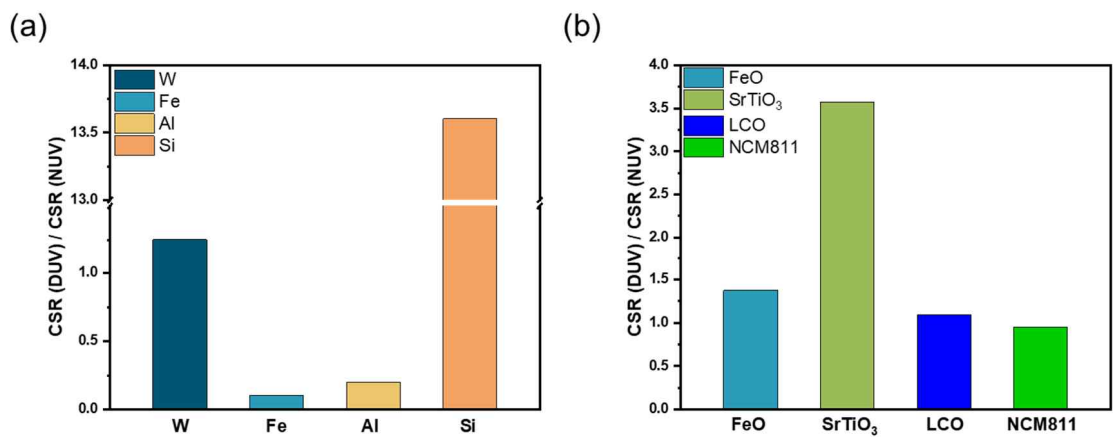

Figure S2. (a) The CSR ratio of DUV to NUV of each element (W, Fe, Al, Si, FeO, SrTiO$_3$, LiCoO$_2$, NCM811)

Table S1. The detection event for NUV and DUV (W, Fe, Al, Si, FeO, SrTiO$_3$, LiCoO$_2$ (LCO), NCM811). The detection event for uncoated LCO and Cr-coated LCO.

| Detection event | W | | Fe | | Al | | Si | |
|---|---|---|---|---|---|---|---|---|
| | DUV | NUV | DUV | NUV | DUV | NUV | DUV | NUV |
| Single | 87.9% | 89.6% | 87.9% | 95.3% | 95.0% | 90.9% | 91.0% | 96.3% |
| Multiple | 7.73% | 9.64% | 11.14% | 3.59% | 3.83% | 8.17% | 7.77% | 2.84% |
| Partial | 4.34% | 0.74% | 0.89% | 1.08% | 1.17% | 0.98% | 1.19% | 0.91% |

| Detection event | FeO | | SrTiO$_3$ | | LiCoO$_2$ | | NCM811 | |
|---|---|---|---|---|---|---|---|---|
| | DUV | NUV | DUV | NUV | DUV | NUV | DUV | NUV |
| Single | 53.7% | 50.7% | 86.4% | 87.2% | 63.6% | 70.4% | 47.9% | 61.4% |
| Multiple | 44.4% | 48.9% | 12.9% | 12.2% | 35.6% | 28.3% | 51.1% | 37.8% |
| Partial | 1.92% | 0.46% | 0.69% | 0.60% | 0.84% | 1.31% | 0.97% | 0.81% |

| Detection event | LiCoO$_2$ (DUV) | |
|---|---|---|
| | Not coated | Coated by Cr |
| Single | 59.7% | 77.7% |
| Multiple | 39.6% | 21.5% |
| Partial | 0.74% | 0.80% |